\documentclass[twocolumn,showpacs]{revtex4}

\usepackage{dcolumn}
\usepackage{amsmath,amssymb}
\ifx\pdfoutput\undefined
  \usepackage[dvips]{graphicx}
\else
  \pdfoutput=1
  \usepackage[pdftex]{graphicx}
\fi

\begin{document}

\title{Single electron transistors with high quality superconducting niobium islands}

\author{R. Dolata}
\email{ralf.dolata@ptb.de}
\author{H. Scherer}
\author{A. B. Zorin}
\author{J. Niemeyer}
\affiliation{Physikalisch-Technische Bundesanstalt, Bundesallee
100, 38116 Braunschweig, Germany}

\date{\today}

\begin{abstract}
Deep submicron Al/AlO$_{\rm x}$/Nb tunnel junctions and single
electron transistors with niobium islands were fabricated by
electron beam gun shadow evaporation. Using stencil masks
consisting of the thermostable polymer polyethersulfone (PES) and
germanium, high quality niobium patterns with good superconducting
properties and a gap energy of up to $2\Delta _{\rm Nb} = 2.5$~meV
were achieved. The $I(U)$ characteristics of the transistors show
special features due to tunneling of single Cooper pairs and
significant gate modulation in both the superconducting and the
normal state.
\end{abstract}


\maketitle

Superconducting quantum coherent devices with small tunnel
junctions offer the possibility of manipulating single Cooper
pairs and may be used to construct current standards
\cite{Likharev85} and charge qubits \cite{Makhlin01}. For optimum
performance, many devices require that the Josephson coupling
energy $E_{\rm J} \equiv (\hbar/2e) \, I_{c} = (R_{\rm
K}/R)\Delta/8$ is of the same order as the charging energy $E_{\rm
c} = e^2/2C_{\rm \Sigma}$, where $R_{\rm K} = h/e^2 \approx
25$~k$\Omega$, $R$ is the junction tunnel resistance, $\Delta$ the
superconducting gap energy, and $C_{\rm \Sigma}$ the total
capacitance of the island. With modern nanotechniques it is
possible to reduce $C_{\rm \Sigma}$ and hence to increase $E_{\rm
c}$ in Al based junctions ($\Delta_{\rm Al} \approx 0.2$~meV), but
it is hard to keep $E_{\rm J} \approx E_{\rm c}$ because this
would require very transparent barriers. With the use of Nb based
junctions ($\Delta_{\rm Nb}^{\rm bulk} \approx 1.5$~meV) it should
be possible to increase the Josephson coupling energy by almost
one order of magnitude for the same tunnel resistance. Further, an
increased $\Delta$ is important since it should suppress
disturbing quasiparticle tunneling onto/from the island
\cite{Tuominen92}. Al-Nb junctions are a first step in this
direction because $2.2\,E_{\rm J}^{\rm Al-Al} \approx E_{\rm
J}^{\rm Al-Nb} \approx 0.3\,E_{\rm J}^{\rm Nb-Nb}$. These
relations follow from the equation for the Josephson coupling
energy in junctions with different electrode materials A and B for
$k_{\rm B}T \ll \Delta_{\rm A},\,\Delta_{\rm B}$ (see, e.g.,
\cite{Barone82}):
\begin{equation*}
E_{\rm J}^{\rm A-B} = \frac{R_{\rm K}}{2\pi R}\,\frac{\Delta_{\rm
A}\cdot\Delta_{\rm B}}{\Delta_{\rm A}+\Delta_{\rm B}}\,{\cal
K}\left(\frac{|\Delta_{\rm A}-\Delta_{\rm B}|}{\Delta_{\rm
A}+\Delta_{\rm B}}\right).
\end{equation*}
Here, ${\cal K}(x)$ is the complete elliptic integral of the first
kind.

It has been shown that rather small Nb based junctions (areas
about 0.1~$\mu$m$^2$) with high quality $I(U)$ characteristics can
be fabricated from Nb/AlO$_{\rm x}$/Nb trilayers by etching and
planarization techniques \cite{Pavolotsky99}, but single charge
devices prepared with these junctions still suffer from high
capacitances ($C_{\rm \Sigma} \approx 6$~fF).

\begin{figure}[h,t]
  \centering
  \includegraphics[scale=0.5,clip=true]{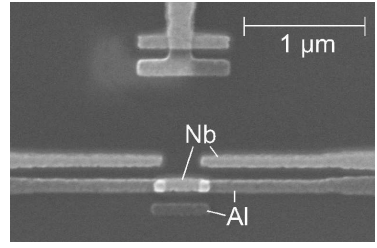}
  \caption{Scanning electron micrograph of an Al/AlO$_{\rm x}$/Nb transistor.
The nominal tunnel junction size is 100~nm $\times$ 100~nm.}
  \label{SEM}
\end{figure}
Shadow or angle evaporation through a suspended mask is a common
technique for the fabrication of ultrasmall tunnel junctions with
low capacitance \cite{Niem-Dolan}. Usually, polymethylmethacrylate
(PMMA) and related copolymers (PMMA-MAA) are used to create high
resolution suspended stencil masks by electron beam lithography.
These masks provide easy lift-off and are successfully used for
the preparation of devices based on soft materials such as
aluminum \cite{Geerligs90}. Unfortunately, for the fabrication of
submicron Nb structures, the use of PMMA entails strong
deterioration of the superconducting properties
\cite{Harada94,Hoss99,Dubos00}. The most important reason for this
is the outgassing of the PMMA polymer during the evaporation of
the high-melting Nb due to radiation heating from the evaporation
source \cite{Hoss99,Dubos00}. For high-melting materials
nonorganic evaporation masks were developed, made of two metallic
\cite{Howard78,Zant90} or two dielectric layers \cite{Hoss99}.
However, the metallic masks pose problems in the fabrication of
deep submicron structures because of their rough edges, and
dielectric masks cannot be lifted off after evaporation. Dubos
{\it et al.} developed a novel lift-off process for the
preparation of high quality submicron Nb structures, based on the
use of the thermostable polymer polyethersulfone (PES)
\cite{Dubos00}. PES remains mechanically tough up to temperatures
of about 200~$^\circ$C and shows very low outgassing compared to
PMMA.

In this paper, we present for the first time results on high
quality superconducting Al/AlO$_{\rm x}$/Nb tunnel junction
samples fabricated using PES based stencil masks. In the first
step, the PES layer (Ultrason E6020P made by BASF, dissolved in
N-methylpyrrolidone (NMP)) was spun onto thermally oxidized
three-inch Si wafers and baked for 20 minutes at 230~$^\circ$C to
form the bottom layer 420~nm in thickness. Next, a 35~nm thick Ge
layer was thermally evaporated onto the PES layer. Finally, a PMMA
layer 130~nm thick was spun on top of the germanium. This layer
was patterned by electron beam lithography and served as a mask
for reactive ion etching of the Ge layer in a CF$_4$ plasma. An
undercut of approximately 300~nm in the PES layer was created by
oxygen plasma etching. In this etching step, the PMMA top layer
was completely removed, the final stencil mask was composed of a
PES bottom layer and a Ge top layer formed the suspended bridges.
The samples were fabricated by two-angle E-gun evaporation in an
ultrahigh vacuum system with a base pressure of about $6 \times
10^{-8}$~mbar onto a substrate wafer thermally anchored to a
copper block at room temperature. The sample was tilted to a first
angle and an Al layer 35~nm in thickness was evaporated. Next, the
AlO$_{\rm x}$ tunnel barrier was formed by oxidation in pure
O$_2$. Process time and pressure were varied for different wafers.
After pumping down of the O$_2$, an Nb layer 40~nm thick was
evaporated using a second tilt angle. Finally, the stencil mask
was lifted off in NMP.

\begin{figure}
  \centering
  \includegraphics[scale=0.3,clip=true]{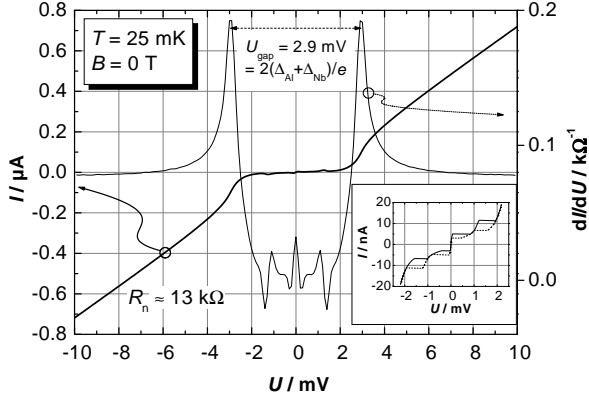}
  \caption{Current-voltage and differential conductivity characteristics for a transistor
   with $R_{\rm n} \approx 13$ k$\Omega$ (lightly oxidized barrier).
   Critical currents of about 5 and 12~nA were observed (see inset).
   According to the maxima of $dI/dU$, the gap voltage was found to be about 2.9~mV,
   corresponding to $\Delta_{\rm Nb}+\Delta_{\rm Al} \approx 1.45$~meV.}
  \label{IVC}
\end{figure}
\begin{figure}
  \centering
  \includegraphics[scale=0.3,clip=true]{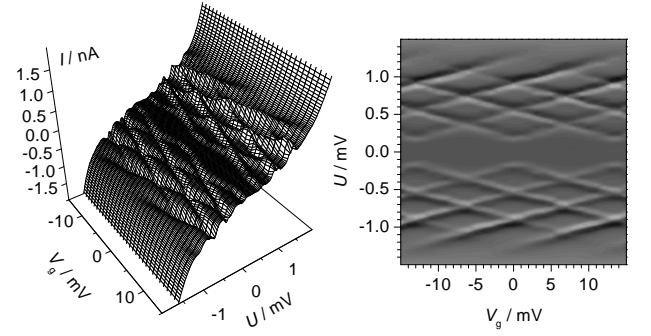}
  \caption{
  Left: 3D plot of the current-voltage characteristics for a
  transistor with a barrier oxidized more strongly
  ($R_{\rm n} = 81$~k$\Omega$, $E_{\rm J} \approx 0.03$~meV~$< E_{\rm c} = 0.15$~meV),
  showing current peak structures in the subgap region ($B = 0$~T).
  Right: Grey-scale plot of the differential conductivity,
  depicting the rhombic structures with a period $\Delta V_{\rm
  g}$ corresponding to $e$, i.e. the period in the normal state achieved at $B=5$~T.}
  \label{3d-a}
\end{figure}
\begin{figure}
  \centering
  \includegraphics[scale=0.3,clip=true]{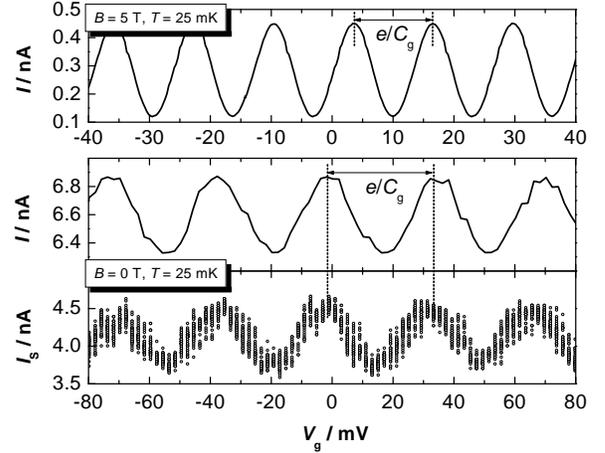}
  \caption{Top: Current modulation of a transistor with $R_{\rm n} =
  81$~k$\Omega$ and $E_{\rm c} = 0.15$~meV, measured in the normal state ($B = 5$~T) at $U = 0.2$~mV.
  Middle: Current modulation of another transistor ($R_{\rm n} = 15$~k$\Omega$, $E_{\rm c} = 0.15$~meV), biased at $U =
  1$~mV~$<\Delta_{\rm Nb}/e$ in the superconducting state ($B = 0$~T). Bottom: Modulation of the switching current $I_{\rm s}$
  of the same sample, recorded while sweeping the bias at $f = 37$~Hz. In all cases, the observed gate modulation was $e$-periodic.}
  \label{Mods}
\end{figure}
We prepared single junctions and single electron transistors that
allowed four terminal measurements to be made. The transistors
consisted of Al wires, an Nb island and an Al/Nb gate electrode
(Fig. \ref{SEM}). Several devices of varying junction size were
included on every wafer, and the maximum nominal junction size was
about 100~nm $\times$ 100~nm. Experiments were carried out in a
dilution refrigerator at a base temperature of $T = 25$~mK.
Electrical biasing and signal amplification were performed using
symmetrical current (or voltage) bias and low-noise preamplifiers
with all signal lines RF filtered by Thermocoax cable pieces 1~m
long. We investigated samples from two wafers prepared under
different barrier oxidation conditions and found a yield of the
working structures of about 80~\%. The sample layouts on the two
wafers were similar except for the length of the island which led
to different gate capacitance values.

On the first wafer (heavier oxidation), we found transistor
samples with gap voltages $U_{\rm gap}=$ between 2.5 and 2.65~mV
according to the maxima of the differential conductivity plots in
the superconducting state ($B= 0$~T). Total resistances $R_{\rm
n}$ from 70 to 250~k$\Omega$ were derived from the $I(U)$-curves
when the samples were driven to the normal state ($B= 5$ T), and
the values for the total island capacitance were $C_{\Sigma}
\approx 0.3$~-~0.6~fF. On the second wafer (lighter oxidation), we
had single Al/AlO$_{\rm x}$/Nb junctions ($U_{\rm gap} \approx
1.45$~meV, $R_{\rm n} = 9$~-~45~k$\Omega$) and transistors with
$U_{\rm gap}= 2.8$~-~2.9~mV, $R_{\rm n} = 12$~-~70~k$\Omega$,
$C_{\Sigma} \approx 0.3$~-~0.6~fF. The $C_{\Sigma}$ achieved in
our Al/Nb transistors thus was of the order of the island
capacitance for typical Al devices \cite{Geerligs90, Tuominen92}.
Switching currents $I_{\rm s}$ up to some nA were measured on
samples of the second wafer (inset in Fig. \ref{IVC})
\cite{Footnote}. The ratios $E_{\rm J}/E_{\rm c}$ of the
transistors were estimated to be in the range of 0.03 - 0.3 for
the first and of 0.16 - 1.2 for the second wafer.

The gap voltages observed are significantly higher compared with
recent results reported for similar Al/AlO$_{\rm x}$/Nb
transistors \cite{Harada94,Born01}. From the maximum gap voltages
of about 2.9~mV observed for single junctions and transistors on
the second wafer, we find that $\Delta_{\rm Nb}+\Delta_{\rm Al}
\approx 1.45$~meV (Fig. \ref{IVC}). Assuming that $\Delta_{\rm Al}
\approx 0.2$~meV \cite{Tuominen92, Harada94, Born01, JQP}, we
derive the gap energy $\Delta_{\rm Nb}$ of the Nb islands to be
about 1.25~meV. Preliminary $I(U)$ measurements showed that
signatures of superconductivity persist up to temperatures of
about $T = 7$ - 8~K, which we identify with the critical
temperature $T_{\rm c}$ of the Nb islands (cf. $T_{\rm c} \approx
9$~K for bulk Nb). According to the formula $2\Delta =
3.52\,k_{\rm B}\,T_{\rm c}$ from BCS theory, this agrees well with
the values for $\Delta_{\rm Nb}$ derived from the gap voltage.

Significant current peaks appeared in the subgap voltage range of
the transistors' $I(U)$ curves, especially marked in samples on
the wafer with the more heavily oxidized barrier where the ratios
$E_{\rm J}/E_{\rm c}$ were only 0.03 - 0.3. These features,
showing up as rhombic structures in the $U$-$V_{\rm g}$ plane
(Fig. \ref{3d-a}), are attributed to different resonant tunneling
processes involving combined single Cooper pair and quasiparticle
tunneling, known as ``Josephson quasiparticle'' cycles \cite{JQP}.
The current peaks could be modulated by the gate voltage with a
period corresponding to $e$. They disappeared when a magnetic
field of $B = 0.3$~T was applied, sufficient to suppress the
superconductivity in the Al electrodes. This supports that the
current peaks were linked with the superconductivity of the two
electrodes. An influence of the magnetic field upon the shapes of
the $I(U)$ curves was observed up to $B \approx 5$~T, indicating
the persistence of superconducting domains in the Nb film island.

Besides the gate modulation in the normal state, we also
investigated the variation of the switching currents $I_{\rm s}$
with $V_{\rm g}$ on the transistors with measurable critical
currents (Fig. \ref{Mods}). Measurements of $I_{\rm s}$ were
performed using a sample-and-hold technique while the sample was
biased with a sawtooth-shaped signal at frequencies $f = 10 -
40$~Hz. In all samples we found that $I_{\rm s}$ was modulated
$e$-periodically. Together with the leakage currents observed in
the subgap regions of the single junctions, this suggests the
presence of quasiparticles that might enter the Nb island via
parasitic conductance channels, or the presence of intra-gap
energy states caused by impurities and allowing quasiparticle
tunneling at low bias.

In summary, we have fabricated Al/AlO$_x$/Nb single electron
tunnel devices by shadow evaporation using a PES based trilayer
mask. This method allows the junction dimensions to be further
reduced. The superconducting Nb islands are characterized by a gap
energy of up to $\Delta_{\rm Nb} = 1.25$~meV close to the bulk
value. Unexpectedly, all gate modulation characteristics measured
on the single electron transistors showed a periodicity
corresponding to $e$, so that further investigations are called
for. The ratio of characteristic energies in these samples varies
over a wide range, clearing the way for devices operating on
single pairs. Our results are a first step towards the highly
desirable Nb/AlO$_x$/Nb junctions for future mesoscopic devices.

The authors wish to thank BASF for providing a sample of PES, and
P. {\AA}gren for discussions about the sample-and-hold measurement
technique.

\end{document}